\documentclass[aps,prd,reprint,superscriptaddress]{revtex4-2}

\usepackage{graphicx,setspace}
\usepackage{dcolumn}
\usepackage{bm,url}
\usepackage[mathlines]{lineno}
\usepackage[linktocpage,colorlinks=true,citecolor=teal,linkcolor=olive,anchorcolor=green,urlcolor=teal]{hyperref}
\usepackage{mathrsfs}
\usepackage{xcolor}
\usepackage{float}
\usepackage[normalem]{ulem}
\usepackage[utf8]{inputenc}
\usepackage{tensor}
\usepackage{physics}
\usepackage{blindtext}
\usepackage{slashed}
\usepackage{tikz}
\usepackage{amsthm,amsfonts,epsfig,natbib,amssymb}
\usepackage{tensor}
\usepackage{mathtools}
\usepackage[scaled]{beramono}
\usepackage[T1]{fontenc}
\usetikzlibrary{decorations.pathmorphing}

\begin{document}

\title{Kinematic topologies of black holes}

\author{Jiayu Yin}
\email{jiayuyin@jxnu.edu.cn}
\affiliation{Department of Physics, Jiangxi Normal University, Nanchang 330022, China}
\author{Jie Jiang}
\email{jiejiang@mail.bnu.edu.cn}
\affiliation{College of Education for the Future, Beijing Normal University, Zhuhai 519087, China}
\author{Ming Zhang}
\email{mingzhang@jxnu.edu.cn, corresponding author }
\affiliation{Department of Physics, Jiangxi Normal University, Nanchang 330022, China}

\begin{abstract}
We investigate the kinematic topologies of light rings (LRs) and massive particle rings (PRs) encircling spherical and axisymmetric black holes. Our results demonstrate that the global topology number of LRs is consistently -1 for asymptotically flat and (Anti-)de Sitter spacetime. Additionally, we show that the global topology of PRs varies, with a value of 0 in asymptotically flat and Anti-de Sitter spacetime but -1 in asymptotic de Sitter spacetime.
\end{abstract}

\maketitle

\section{Introduction}

Recent observations, including gravitational wave measurements from coalescing binaries \cite{LIGOScientific:2016aoc} and images of supermassive black hole candidates in M87 \cite{EventHorizonTelescope:2019dse} and Sgr A* \cite{EventHorizonTelescope:2022wkp} by the Event Horizon Telescope, have allowed for the identification of black holes in our universe. In addition to predicting the existence of an event horizon \cite{Penrose:1969pc}, general relativity also describes the behavior of surrounding matter, which can be modeled as massless photons and massive particles. By studying the kinematics of these particles, researchers can identify the most salient geodesic features, including null-like light rings (LRs) of photons and timelike circular orbits of massive particles, referred to as particle rings (PRs), around central black holes.

The event horizon is a global property of a black hole spacetime and does not form at a particular time \cite{Cardoso:2019rvt}. Similarly, null-like LRs do not evolve with time \cite{Claudel:2000yi}. However, the existence and properties of PRs depend on the energy and angular momentum of the massive particles. A recent study using topology \cite{Cunha:2020azh} has proven that an equilibrium stationary, axisymmetric, asymptotically flat black hole spacetime will always have at least one LR. This conclusion also holds for horizonless spacetimes \cite{Guo:2020qwk} and asymptotically Anti-de Sitter (AdS) \cite{Ghosh:2021txu} and de Sitter (dS) spherical spacetimes \cite{Wei:2020rbh}. Researchers have also investigated the kinematic topologies of PRs around asymptotically flat, spherically symmetric black holes \cite{Ye:2023gmk} and axisymmetric black holes \cite{Wei:2022mzv}, discovering that PRs always come in pairs for a fixed angular momentum of the massive particles. Besides, we note that it was shown that the asymptotically Melvin spacetime may admit no LRs outside the event horizon \cite{Junior:2021dyw}.

Black holes are surrounded by massless photons and massive particles, which are essential to astrophysical observations. A question that arises is whether an equilibrium black hole spacetime always has null-like LRs and PRs in a general setting. This paper aims to answer this intriguing question by generalizing the findings of previous studies \cite{Cunha:2020azh,Wei:2020rbh,Ye:2023gmk,Wei:2022mzv}. Our proposed theorem on the kinematic topologies of LRs and PRs around black holes is as follows: A (3+1)-dimensional, non-extremal, stationary, axisymmetric (or static spherically symmetric) black hole spacetime with $\mathbb{Z}_2$ symmetry and a spherical Killing horizon, will have an odd number of LRs, irrespective of the spacetime's asymptotic properties. In contrast, the number of PRs is even for asymptotically flat and asymptotically AdS spacetimes but odd for asymptotically dS spacetimes.

While null-like LRs correspond to distinct radial orbits, PRs are a continuum of connected orbits with varying angular momentum and energy \cite{Wei:2022mzv,Ye:2023gmk}. However, based on these studies, we can establish topological properties for PRs with a given angular momentum. Additionally, marginally stable circular orbits (MSCOs) exist for massive particles \cite{Cunha:2022nyw}, which are determined solely by black hole properties and have unique angular momenta and energy. Two primary types of MSCOs are the innermost stable circular orbits (ISCOs) \cite{Bardeen:1972fi} and the outermost stable circular orbits (OSCOs) \cite{Boonserm:2019nqq}. We will analyze the topological properties of these critical orbits as bifurcation points.

\section{\label{ssbh} LRs and PRs around spherically symmetric black holes}
We will now investigate the LRs and PRs in a spherically symmetric black hole spacetime background. We consider all three spacetime asymptotic structures: flat, AdS, and dS. The exterior region of  a static, spherically symmetric spacetime can be described by the metric
\begin{equation}\label{geo}
ds^2=f(z)dt^2-\frac{dz^2}{g(z)}-z^2 (d\theta^2 +\sin^{2}\theta d\phi^2),
\end{equation}
where $f(z)$ and $g(z)$ are functions of the radial coordinate $z$. The event horizon $r_h$ of the black hole is the largest root of $g(z)=0$. We suppose the black hole owns $\mathbb{Z}_2$ symmetric property, such that the circular orbit of the massless or massive particle will be on the equatorial plane.  The effective potential of the particle around the black hole (\ref{geo}) is \cite{Chandrasekhar:1985kt}
\begin{equation}\label{effepo}
V_e =g(z) \left(\epsilon+\frac{L^2}{z^2}-\frac{E^2}{f(z)}\right),
\end{equation}
where we have denoted the particle's specific energy as $E$ and angular momentum as $L$. We have the dimensionless quantity $\epsilon=0, 1$ for the photons and massive particles, respectively. The radial position $z_i$ of the particle on the circular orbit depends exclusively on the radial effective potential and can be obtained by demanding that
$V_e (z_i)=0, V_e^\prime (z_i)=0$, which are equivalent to the zero points of a characterized function
\begin{equation}\label{bbV}
\mathbb{V}(z)\equiv\partial_z E(z)=\frac{L^2 z f'(z)+z^3 \epsilon  f'(z)-2 L^2 f(z)}{2 z^2 \sqrt{f(z) \left(L^2+z^2 \epsilon \right)}},
\end{equation}
where $E(z)= \sqrt{\left(\epsilon+L^2/z^2\right) f(z)}$. We can denote 
$\mathcal{P}(z)\equiv 2 z^2 \sqrt{f(z) \left(L^2+z^2 \epsilon \right)}, \,\mathcal{S}(z)\equiv L^2 z f'(z)+z^3 \epsilon  f'(z)-2 L^2 f(z)$.
To determine the roots of the equation $\mathbb{V}(z)=0$, we can analyze the asymptotic behavior of the characterized function $\mathbb{V}(z)$. First, it is evident that $\mathcal{P}(z)>0,\quad \text{for}\quad z>r_h$, and $\mathcal{P}(z)\to 0^+$ for $z\to r_h^+$. So the sign of the function $\mathbb{V}(z)$ near the horizon is totally determined by $\mathcal{S}(z)$.  When approaching the event horizon of the black hole, we have
$
\left.\mathcal{S}(z)\right|_{z\to r_h}=\epsilon r_h^3 f^{\prime}(r_h)+L^2 r_h f^{\prime}(r_h)>0,
$
regardless of the kind of the particle and the asymptotic properties of the spacetime. Here we have used the condition that  the surface gravity $\kappa$ of the nonextremal spherically symmetric  black hole (\ref{geo}) is positive:
$
\kappa= \lim _{z \rightarrow r_h} \sqrt{g(z)/f(z)} f^{\prime}(z)/2>0.
$

We now need to analyze the asymptic behavior of the characterized function $\mathbb{V}(z)$ in the spacetime boundary. First, it is straightforward that
$
\mathcal{P}(z)>0,\quad \text{for}\quad z>r_b,
$
where $r_b$ denotes the spacetime boundary. In the asymptotically AdS and flat spacetimes, we have $r_b\to \infty$; in the asymptotically dS spacetime, we have $r_b=r_c$ with $r_c$ the cosmological horizon of the black hole spacetime.

To analyze the behavior of the function $\mathcal{S}(z)$ at the spatial boundary, we should consider different asymptotic properties of the spacetime. The asymptotic expressions for the blackening factor at the spatial boundary are different for flat, AdS, and dS black hole spacetime, as
\begin{equation}
f(z)\to\left\{\begin{array}{l}
1-\frac{m}{z}+\mathcal{O}\left(\frac{1}{z^2}\right), \text{flat;}\\
1-\frac{m}{z}-\frac{\Lambda z^2}{3}+\mathcal{O}\left(\frac{1}{z^2}\right), \Lambda<0, \text{AdS;} \\
1-\frac{m}{z}-\frac{\Lambda z^2}{3}+\mathcal{O}\left(\frac{1}{z^2}\right), \Lambda>0, \text{dS.}
\end{array}\right.
\end{equation}
The cosmological constant $\Lambda$ is related to the curvature radius of the spacetime as $l$ by the relation $|\Lambda |=3/l^2$ and $m$ is the mass parameter. Note that $|m^2\Lambda |\ll  1$. Then according to the above asymptotic expressions of the blackening factor of the black hole, we have
\begin{equation}\label{szfar}
\mathcal{S}(z)\to\left\{\begin{array}{l}
-2L^2+mz\epsilon+\mathcal{O}\left(\frac{1}{z}\right), \text{flat; }\\
-2L^2+mz\epsilon-\frac{2}{3}\Lambda  z^4 \epsilon+\mathcal{O}\left(\frac{1}{z}\right), \text{AdS;} \\
-2L^2+mz\epsilon-\frac{2}{3}\Lambda  z^4 \epsilon+\mathcal{O}\left(\frac{1}{z}\right), \text{dS.}
\end{array}\right.
\end{equation}
Obviously, for the photon ($\epsilon=0$), we have
$
\left.\mathcal{S}(z)\right|_{z\to r_b}<0
$
for all three cases. For the massive particle ($\epsilon=1$), we have
\begin{equation}
\left.\mathcal{S}(z)\right|_{z\to r_b}\to\left\{\begin{array}{l}
mz >0, \text{flat;}\\
mz-\frac{2}{3}\Lambda  z^4 \epsilon >0, \text{AdS;}\\
-\frac{2}{3}\Lambda  z^4 \epsilon<0, \text{for dS}
\end{array}\right.
\end{equation}
for any given angular momentum $L$ of the particle.

\begin{table*}[ht!]
\caption{\label{tab1}Behavior of the characterized function $\mathbb{V}(z)$ at the boundaries of the spacetime.}
    \centering
    \renewcommand\arraystretch{1.8}
\begin{tabular}{ccccc}
\hline  & flat & AdS & dS  \\
 \hline
    massless photon& $\mathbb{V}(r_h)>0, \mathbb{V}(r_b)<0$ & $\mathbb{V}(r_h)>0, \mathbb{V}(r_b)<0$ & $\mathbb{V}(r_h)>0, \mathbb{V}(r_b)<0$ \\

    massive particle & $\mathbb{V}(r_h)>0, \mathbb{V}(r_b)>0$ & $\mathbb{V}(r_h)>0, \mathbb{V}(r_b)>0$ & $\mathbb{V}(r_h)>0, \mathbb{V}(r_b)<0$ \\
\hline
\end{tabular}
\end{table*}

As a result, the asymptotic behaviors of the characterized function for the massless photons and massive particles around the spherically symmetric black holes can be collected in Table \ref{tab1}. We can see that the characterized function $\mathbb{V}(z)$ for the massless photon share the same asymptotic behaviors in asymptotically flat, AdS, and dS spacetimes, and it is the same for that of the massive particle in the dS spacetime. It means that there are odd number of LRs in all three kinds of spacetime and odd number of  PRs for each given angular momentum of the massive particle in the dS spacetime. However, we can find that the asymptotic behaviors of the massive particles are different in the asymptotically flat and AdS spacetimes, so that we can infer that there are even number of PRs for each given angular momentum of the massive particles.

\section{\label{asbh} LRs and PRs around axisymmetric black holes}
Now we generalize the above analysis of the behaviors of the LRs and PRs in the spherically  symmetric black hole spacetime to the axisymmetric black hole spacetime case which (i) has Killing vectors $(\partial_t)^a$ and $(\partial_\varphi)^a$, (ii) is circular and at least $C^2$-smooth on and outside the  nonextremal, topologically spherical Killing horizon. We set the metric of the axisymmetric black hole with $\mathbb{Z}_2$ symmetry and a Lorenzian $(-,+,+,+)$ signature as 
\begin{equation}
\begin{aligned}
d s^2=&g_{t t}(z, \theta) d t^2+g_{z z}(z, \theta) d z^2+g_{\theta \theta}(z, \theta) d \theta^2\\&+g_{\varphi \varphi}(z,\theta) d \varphi^2+2 g_{t \varphi}(z,\theta) d t d \varphi.
\end{aligned}
\end{equation}
The $\mathbb{Z}_2$ symmetry ensures that the circular orbits are on the equatorial plane. Outside the event horizon $r_h$ and in the domain of outer communication of the black hole, we have $g_{z z}>0, g_{\theta \theta}>0, g_{\varphi \varphi}>0$. We have chosen the gauge that the polar coordinate $\theta$ is orthogonal to the radial coordinate $z$, such that $g_{z\theta}=0$. The ranges of the coordinates are $t \in(-\infty,+\infty), \theta \in[0, \pi], \varphi \in[0,2 \pi)$. We here will consider the asymptotically flat, AdS, and dS spacetime, so $z \in\left[r_H, r_b)\right.$, where $r_b\to \infty$ for the former two cases and $r_b\to r_c$ with $r_c$ the cosmological horizon for the last case. Both $r_h$ and $r_c$ are Killing horizons $\mathcal{H}$, which are given by $\left.\left(\chi^\mu \chi_\mu\right)\right|_{\mathcal{H}}=0$, with $\chi=\partial_t+\omega_H \partial_{\varphi}$ and $\omega_H=-\left.\left(g_{t \varphi} / g_{\varphi \varphi}\right)\right|_{\mathcal{H}}$.
The determinant of this metric is
$
\mathrm{Det}(-g)=g_{\theta \theta } g_{zz} \left(g_{t\varphi}^2-g_{tt} g_{\varphi \varphi }\right),
$
where, in the domain of outer communication, we have  $B(z, \theta) \equiv g_{t \varphi}^2-g_{t t} g_{\varphi \varphi}>0$ and on the Killing horizon, $B(z, \theta)=0$ \cite{Cunha:2020azh}.

The radial effective potential of the massive particle in this spacetime can be written as \cite{Delgado:2021jxd,Wei:2022mzv}
$
V_e =-\frac{g_{\varphi \varphi}}{B}\left(E-e\right)\left(E-\tilde{e}\right),
$
where
$
e=\left(-L g_{t \varphi}+\sqrt{B} \sqrt{L^2+\epsilon^2 g_{\varphi \varphi}}\right)/g_{\varphi \varphi},
$
$
\tilde{e}=\left(-L g_{t \varphi}-\sqrt{B} \sqrt{L^2+\epsilon^2 g_{\varphi \varphi}}\right)/g_{\varphi \varphi}.
$
In the absence of ergoregions, $\tilde{e}$ is negative, so it is sufficient for us to consider only $e$ \cite{Collodel:2017end}. The circular orbit of the particle can be obtained by 
$
V_e (z)=0 \Rightarrow E=e, V_e^{\prime}(z)=0 \Rightarrow \partial_z e(z)=0.
$
Note that for the stable (unstable) circular orbits, we have $V_e^{\prime\prime}>0\, (<0)$; for the MSCOs, we have $V_e^{\prime\prime}=0$.
As a result, we can define a similiar function as (\ref{bbV}),
\begin{equation}\label{bbV2}
\mathbb{V}(z)\equiv\partial_z e(z)=\frac{\mathcal{S}(z)}{\mathcal{P}(z)},
\end{equation}
where $$\mathcal{P}(z)=2g_{\varphi\varphi}^2 \tilde{l}\sqrt{B}>0,$$ 
$$
\begin{aligned}
\mathcal{S}(z)=&-\epsilon^2 g_{\varphi \varphi}^3 g_{t t}^{\prime}+g_{\varphi \varphi}^2\left(-L^2 g_{t t}^{\prime}+2 \epsilon^2 g_{t \varphi} g_{t \varphi}^{\prime}\right)\\&+2 L g_{t \varphi}\left(-L g_{t \varphi}+\tilde{l} \sqrt{B}\right) g_{\varphi \varphi}^{\prime}+2g_{\varphi \varphi} L^2 g_{t \varphi} g_{t \varphi}^{\prime}\\&+g_{\varphi \varphi}\left[\left(L^2 g_{t t}-\epsilon^2 g_{t \varphi}^2\right) g_{\varphi \varphi}^{\prime}-2 L \tilde{l} \sqrt{B} g_{t \varphi}^{\prime}\right],
\end{aligned}
$$
with $\tilde{l}=\sqrt{L^2+\epsilon^2 g_{\varphi\varphi}}$. The primes denote differentiation with respect to $r$.

The roots of $\mathbb{V}(z)=0$ gives the radii of the LRs and PRs. To determine the existence of the PRs, we need to analyze the asymptotic behavior of the characterized function $\mathbb{V}(z)$ at the boundary of the domain of outer communication of the black hole spacetime.

First, let us consider the asymptotically flat case. In this case, at the event horizon, we have 
$
g_{t \varphi}=-\omega_H g_{\varphi \varphi},
$
where $\omega_H=\left.\left(g_{t \varphi} / g_{\varphi \varphi}\right)\right|_{z\to r_h}$ is dependent only on the black hole's parameters. Then we obtain
$
\left.\mathcal{S}(z)\right|_{z\to r_h}= g_{\varphi \varphi}\left[L^2 g_{t t} g_{\varphi \varphi}^{\prime}-L^2 g_{\varphi \varphi} g_{t t}^{\prime}+\epsilon^2 g_{\varphi \varphi}^2\left(\omega_H^2 g_{\varphi \varphi}^{\prime}-g_{t t}^{\prime}\right)\right].
$
Considering that  $B=0$ at the horizon, we further get
$
g_{tt}(r_h)=g^2_{t\varphi}(r_h)/g_{\varphi\varphi}(r_h)>0.
$
Derivative with respect to $r_h$ then yields
$
g_{\varphi \varphi}^2\left(r_h\right) g_{t t}^{\prime}\left(r_h\right)=2 g_{t \varphi}\left(r_h\right) g_{t \varphi}{ }^{\prime}\left(r_h\right) g_{\varphi \varphi}\left(r_h\right)-g_{t \varphi}^2\left(r_h\right) g_{\varphi \varphi}^{\prime}\left(r_h\right)<0,
$
where we have used general assumptions that should be satisfied by the axisymmetric black holes as 
$
 g_{t \varphi}\left(r_h\right)g_{t \varphi}^{\prime}\left(r_h\right)<0, 
g_{\varphi \varphi}\left(r_h\right)>0, g_{\varphi \varphi}^{\prime}\left(r_h\right)>0.
$
Thus we have 
$
\left.\mathcal{S}(z)\right|_{z\to r_h}>0
$
for both $\epsilon=0, 1$ cases. At spatial infinity, we have
\begin{equation}
g_{tt}\to\left\{\begin{array}{l}
-1, \text{flat;}\\
-1+\frac{z^2 \Lambda}{3}, \Lambda<0, \text{AdS;} \\
-1+\frac{z^2 \Lambda}{3}, \Lambda>0, \text{dS,}
\end{array}\right.
\end{equation}
and $g_{\varphi \varphi} \rightarrow z^2$. Then for the asymptotically flat case, the spacetime  reaches flat spacetime in standard spherical coordinates, so the conclusion is the same as that for the spherical case as derived above.

For the asymptotically AdS case, at the spatial infinity, using the relation
$
g_{t \varphi}=-\omega_{\Lambda<0} g_{\varphi \varphi},
$
where $\omega_{\Lambda<0}=\left.\left(g_{t \varphi} / g_{\varphi \varphi}\right)\right|_{z\to \infty}$, we have
$
\left.\mathcal{S}(z)\right|_{z\to \infty}=-2 L^2 z^3-\frac{2}{3} \epsilon^2 z^7 \left(\Lambda -3 \omega_{\Lambda<0}^2\right).
$
Then it is evident that for the massless photon, we have $\left.\mathcal{S}(z)\right|_{z\to \infty}<0$ and for the massive particle, we have $\left.\mathcal{S}(z)\right|_{z\to \infty}>0$.

For the asymptotically dS case, using the relation
$
\lim_{r\to r_c}B(z, \theta) \to 0,
$
where $r_c$ is the cosmological horizon, which is also a Killing horizon, we have
$
\left.\mathcal{S}(z)\right|_{z\to r_c}=2 \left(\sqrt{2}-2\right) L z^3<0
$
for the massless particle and
$
\left.\mathcal{S}(z)\right|_{z\to r_c}=6 \sqrt{3} \left(1/\Lambda \right)^{\frac{5}{2}} \left(\Lambda  L \sqrt{3/\Lambda+L^2}-\Lambda  L^2-3\right)<0
$
for the massive particle. Note that for the last case, the necessary and sufficient condition is $\Lambda>0,\, L\in \mathbb{R}$.

To sum up,  the asymptotic properties of the characterized function for the LRs and PRs  around the axisymmetric black hole  are the same with those for the spherical black holes as shown in Table \ref{tab1}.

\section{\label{top}Topological categories}
We have analyzed the asymptotic property of the characterized function $\mathbb{V}(z)$ whose zero points determine the location of the LRs and PRs around the spherically symmetric and axisymmetric black holes. We will now convert the asymptotic properties of $\mathbb{V}(z)$ to the topological description, i.e., we will endow topological properties to the LRs and PRs (including the MSCOs)  around the black holes. 

To this end, we first try to define the topological number of the LRs and PRs, which we name as residue method. This method is different to the topological current method raised in \cite{Ye:2023gmk,Wei:2022mzv}. Let's consider a complex function $\mathcal{C}(z)$ that's analytic on the complex plane except for a few isolated points $z_1, z_2, \ldots, z_n$ that are poles of the first order. We can assign a real topology number, denoted as $\omega_k$, to one of these isolated points, say, $z_k$. The topology number of $\mathcal{C}(z)$ at $z_k$ can be defined as the sign function $\text{Sgn}$ of the residue $\text{Res} \mathcal{C} (z_k)$, as
\begin{equation}
\omega_k \equiv \text{Sgn}[\text{Res} \mathcal{C} (z_k)]=\frac{\text{Res} \mathcal{C} (z_k)}{\left|\text{Res} \mathcal{C} (z_k)\right|}.
\end{equation} 
By using these topology numbers, we can reflect the local topological property of the function $\mathcal{C}(z)$. Additionally, we can define the global topological number $W$ of the function as the sum of all the topology numbers $\omega_k$ as
\begin{equation}
W\equiv\sum_k \omega_k.
\end{equation}

\begin{table*}[ht!]
\caption{\label{tab2} Local topological numbers (LTNs) and global topological numbers (GTNs) of LRs and PRs in asymptotically flat, dS, and AdS black hole spacetimes. We also list existing literature results we have recovered and what we have newly found (marked with $\cross$ sign).}
    \centering
    \renewcommand\arraystretch{1.8}
\begin{tabular}{ccccc}
\hline  & LR in flat, dS, AdS & PR in dS & PR in flat, AdS  \\
 \hline
    LTN & $-1$ (unstable), $1$ (stable) & $-1$ (unstable), $1$ (stable) & $-1$ (unstable), $1$ (stable) \\

    GTN  & -1 & -1 & 0 \\
    Spherical & flat \cite{Wei:2020rbh}, dS  \cite{Wei:2020rbh,Ghosh:2021txu}, AdS  \cite{Wei:2020rbh,Ghosh:2021txu}   & [$\cross$] & flat \cite{Ye:2023gmk,Cunha:2022nyw}, AdS [$\cross$]\\
    Axial & flat \cite{Cunha:2020azh,Guo:2020qwk}, dS [$\cross$],  AdS [$\cross$] & [$\cross$] &flat  \cite{Wei:2022mzv}, AdS [$\cross$]\\
\hline
\end{tabular}
\end{table*}

According to our former definition of the characterized function $\mathbb{V}(z)$ in \eqref{bbV} and \eqref{bbV2}, we can specifically define
$
\mathcal{C}(z)\equiv \mathbb{V}^{-1}(z)=\mathcal{P}(z)/\mathcal{S}(z)\equiv \partial_z^{-1} e(z),
$
where $\mathcal{P}(z)>0$ and $\mathcal{S}(z)$ are regular polynomials without singular points as shown above. Then, for the LRs around the black holes, based on the asymptotic property of the characterized function $\mathcal{C}(z)$ at the  horizon and spatial boundary, we can infer that there must be odd number of real roots for  $\mathcal{S}(z)=0$, which we denote as $r_h<r_1<r_2<r_3\cdots<r_{2k-1}<r_b$, with $1\leqslant k\in \mathbb{N}$. Thus we can recast the complex function $\mathcal{S}(z)$ as
$
\mathcal{S}(z)=\bar{\mathcal{A}}(z)\left(r_1-z\right)\left(z-r_2\right) \cdots\left(z-r_{2 k-2}\right)\left(r_{2 k-1}-z\right),
$
where $\bar{\mathcal{A}}(z)$ is a regular polynomial that does not have real roots. 
Then we have
\begin{align}\label{tonu}
\omega_{2k}&=\text{Sgn}(\text{Res}(\mathcal{C}(z_{2k})))=1,\\
 \omega_{2k-1}&=\text{Sgn}(\text{Res}(\mathcal{C}(z_{2k-1})))=-1.
\end{align}
The topological numbers can reflect the stability of the LRs. According to our definition of the characterized function $\mathbb{V}(z)$, the LRs are stable (unstable) iff $\mathbb{V}^{\prime}(z)>0\, (<0)$. The topological number $-1, 1$  correspond to unstable LRs and stable LRs, respectively.
The total topological charge of the LRs should be
\begin{equation}
W_{\mathrm{LR}}=\sum_{i=1}^{2k-1}\omega_i =-1.
\end{equation}
Thus we can conclude that there must be at least one LR outside the event horizon of the black  hole. This is independent of the asymptotic properties of the spacetime.

For the PRs, as shown in Table \ref{tab1}, we see there are two categories of asymptotic properties, one is for the  asymptotic dS case, which is the same with the LRs cases, and the other is for the asymptotic flat and AdS cases. For the PRs of the massive particles around the asymptotically dS black holes, based on the asymptotic property of the characterized function $\mathcal{C}$ at horizon and spatial boundary in the dS spacetime, we can directly know that the total topological charge should be 
\begin{equation}
W_{\mathrm{PR}}^{\mathrm{dS}}=-1,
\end{equation}
which means that the number of  PRs for a given angular momentum of the massive particle should be odd. This has been verified by the Kerr-dS case, where there can be one or three circular orbits for  certain value of particle's angular momentum.

For the PRs of the massive particles around the asymptotically AdS and flat black holes, according to the asymptotical property of the characterized function $\mathcal{C}(z)$ at horizon and spatial boundary, we know that there must be even number of real roots for $\mathcal{S}(z)=0$, for which we can denote as $r_h<r_1<r_2<r_3\cdots<r_{2k}<r_b$, and can write the function $\mathcal{S}(z)$ as
$
\mathcal{S}(z)=\bar{\mathcal{B}}(z)\left(r_1-z\right)\left(z-r_2\right) \cdots \left(r_{2 k-1}-z\right)\left(z-r_{2 k}\right),
$
where $\bar{\mathcal{B}}(z)$ is a regular polynomial that does not have real roots. As a result, we get $\omega_{2k}=1,\,  \omega_{2k-1}=-1$. This result is consistent with \eqref{tonu}. We can also know that the topological number $-1, 1$  correspond to unstable PRs and stable PRs, respectively.
The total topological charge of the PRs in the asymptotic flat and AdS spacetime should be 
\begin{equation}
W_{\mathrm{PR}}^{\mathrm{flat, AdS}}=\sum_{i=1}^{2k}\omega_i =0.
\end{equation}
This result implies that for a given angular momentum of the massive particle around the asymptotically flat or AdS black hole, there should be even number of PRs outside the event horizon.

We now would like to consider the bifurcation points, which correspond to the MSCOs of the massive particles. The ISCOs exist in all three kinds of spacetimes with different asymptotic properties. However, the OSCOs exist only in the asymptotic dS spacetime. Algebraically, an MSCO corresponds to a second order pole point, say, $r_m$, of the characterized function $\mathcal{C}(z)$. $r_m$ can be viewed as a merging point of two adjacent points $r_{2k-1}$ and $r_{2k}$, which are pole points of the first order. As a result, the topological number of an MSCO can be obtained as
\begin{equation}
\omega_m=\lim_{r_{2k-1}\to r_m \atop r_{2k}\to r_m}(\omega_{2k-1}+\omega_{2k})=0.
\end{equation}
One can see that this respects the conservation of the total topological number for the PRs.

\section{Concluding remarks}
We conclude the results of local topological numbers and global topological numbers for the LRs and PRs in the  asymptotically flat, dS, and AdS black hole spacetime in Table \ref{tab2}. On the one hand, we can see that the local topological number can reflect the local stability of the particle's circular orbits, and the stable (unstable) LRs and PRs have topological numbers $\mp 1$. On the other hand, we can see from the table that the global topological number can reflect the number of the LRs and PRs. Our result implies that there are odd number of (at least one) LRs around the asymptotically flat, dS, and AdS black holes and odd number of PRs for a given angular momentum of the massive particle in the asymptotically dS black hole; in contrast, there are even number of (maybe zero) PRs for a given angular momentum of the massive particle around the asymptotically flat and AdS black holes. Moreover, the topological number of the MSCOs for the PR is $0$, which obeys the topological number conservation.

In Table \ref{tab2}, we summarize the results of our analysis of local and global topological numbers for null-like LRs and PRs in asymptotically flat, dS, and AdS black hole spacetimes. We find that the local topological number characterizes the stability of the particle's circular orbits, with stable (unstable) LRs and PRs having topological numbers of $\mp 1$, respectively. Meanwhile, the global topological number reflects the number of LRs and PRs present. Our analysis shows that there are always an odd number of (at least one) LRs around asymptotically flat, dS, and AdS black holes, and an odd number of PRs for a given angular momentum of the massive particle in asymptotically dS black holes. In contrast, around asymptotically flat and AdS black holes, there are an even number of (maybe zero) PRs for a given angular momentum. Additionally, the topological number of the marginally stable circular orbits (MSCOs) for the PRs is zero, which follows the topological number conservation.

Lastly, we would like to emphasize that we have recovered all former results in existing literature, as well as complement all cases in the asymptotically dS and AdS spacetime, as was shown in Table \ref{tab2}.  The residue method we have used is more efficient and easier than existing methods that need to specify the asymptotic behaviors of the effective potentials in the angular direction in literature.

\begin{acknowledgments}
We would like to thank Shao-Wen Wei for useful discussions. MZ is supported by the National Natural Science Foundation of China with Grant No. 12005080.  JJ  is supported by the National Natural Science Foundation of China with Grant No.  210510101, the Guangdong Basic and Applied Research Foundation with Grant No. 217200003, and the Talents Introduction Foundation of Beijing Normal University with Grant No. 310432102.
\end{acknowledgments}


\begin{thebibliography}{19}%
\makeatletter
\providecommand \@ifxundefined [1]{%
 \@ifx{#1\undefined}
}%
\providecommand \@ifnum [1]{%
 \ifnum #1\expandafter \@firstoftwo
 \else \expandafter \@secondoftwo
 \fi
}%
\providecommand \@ifx [1]{%
 \ifx #1\expandafter \@firstoftwo
 \else \expandafter \@secondoftwo
 \fi
}%
\providecommand \natexlab [1]{#1}%
\providecommand \enquote  [1]{``#1''}%
\providecommand \bibnamefont  [1]{#1}%
\providecommand \bibfnamefont [1]{#1}%
\providecommand \citenamefont [1]{#1}%
\providecommand \href@noop [0]{\@secondoftwo}%
\providecommand \href [0]{\begingroup \@sanitize@url \@href}%
\providecommand \@href[1]{\@@startlink{#1}\@@href}%
\providecommand \@@href[1]{\endgroup#1\@@endlink}%
\providecommand \@sanitize@url [0]{\catcode `\\12\catcode `\$12\catcode
  `\&12\catcode `\#12\catcode `\^12\catcode `\_12\catcode `\%12\relax}%
\providecommand \@@startlink[1]{}%
\providecommand \@@endlink[0]{}%
\providecommand \url  [0]{\begingroup\@sanitize@url \@url }%
\providecommand \@url [1]{\endgroup\@href {#1}{\urlprefix }}%
\providecommand \urlprefix  [0]{URL }%
\providecommand \Eprint [0]{\href }%
\providecommand \doibase [0]{https://doi.org/}%
\providecommand \selectlanguage [0]{\@gobble}%
\providecommand \bibinfo  [0]{\@secondoftwo}%
\providecommand \bibfield  [0]{\@secondoftwo}%
\providecommand \translation [1]{[#1]}%
\providecommand \BibitemOpen [0]{}%
\providecommand \bibitemStop [0]{}%
\providecommand \bibitemNoStop [0]{.\EOS\space}%
\providecommand \EOS [0]{\spacefactor3000\relax}%
\providecommand \BibitemShut  [1]{\csname bibitem#1\endcsname}%
\let\auto@bib@innerbib\@empty
%</preamble>
\bibitem [{\citenamefont {Abbott}\ \emph {et~al.}(2016)\citenamefont {Abbott}
  \emph {et~al.}}]{LIGOScientific:2016aoc}%
  \BibitemOpen
  \bibfield  {author} {\bibinfo {author} {\bibfnamefont {B.~P.}\ \bibnamefont
  {Abbott}} \emph {et~al.} (\bibinfo {collaboration} {LIGO Scientific,
  Virgo}),\ }\href {https://doi.org/10.1103/PhysRevLett.116.061102} {\bibfield
  {journal} {\bibinfo  {journal} {Phys. Rev. Lett.}\ }\textbf {\bibinfo
  {volume} {116}},\ \bibinfo {pages} {061102} (\bibinfo {year} {2016})},\
  \Eprint {https://arxiv.org/abs/1602.03837} {arXiv:1602.03837 [gr-qc]}
  \BibitemShut {NoStop}%
\bibitem [{\citenamefont {Akiyama}\ \emph {et~al.}(2019)\citenamefont {Akiyama}
  \emph {et~al.}}]{EventHorizonTelescope:2019dse}%
  \BibitemOpen
  \bibfield  {author} {\bibinfo {author} {\bibfnamefont {K.}~\bibnamefont
  {Akiyama}} \emph {et~al.} (\bibinfo {collaboration} {Event Horizon
  Telescope}),\ }\href {https://doi.org/10.3847/2041-8213/ab0ec7} {\bibfield
  {journal} {\bibinfo  {journal} {Astrophys. J. Lett.}\ }\textbf {\bibinfo
  {volume} {875}},\ \bibinfo {pages} {L1} (\bibinfo {year} {2019})},\ \Eprint
  {https://arxiv.org/abs/1906.11238} {arXiv:1906.11238 [astro-ph.GA]}
  \BibitemShut {NoStop}%
\bibitem [{\citenamefont {Akiyama}\ \emph {et~al.}(2022)\citenamefont {Akiyama}
  \emph {et~al.}}]{EventHorizonTelescope:2022wkp}%
  \BibitemOpen
  \bibfield  {author} {\bibinfo {author} {\bibfnamefont {K.}~\bibnamefont
  {Akiyama}} \emph {et~al.} (\bibinfo {collaboration} {Event Horizon
  Telescope}),\ }\href {https://doi.org/10.3847/2041-8213/ac6674} {\bibfield
  {journal} {\bibinfo  {journal} {Astrophys. J. Lett.}\ }\textbf {\bibinfo
  {volume} {930}},\ \bibinfo {pages} {L12} (\bibinfo {year}
  {2022})}\BibitemShut {NoStop}%
\bibitem [{\citenamefont {Penrose}(1969)}]{Penrose:1969pc}%
  \BibitemOpen
  \bibfield  {author} {\bibinfo {author} {\bibfnamefont {R.}~\bibnamefont
  {Penrose}},\ }\href {https://doi.org/10.1023/A:1016578408204} {\bibfield
  {journal} {\bibinfo  {journal} {Riv. Nuovo Cim.}\ }\textbf {\bibinfo {volume}
  {1}},\ \bibinfo {pages} {252} (\bibinfo {year} {1969})}\BibitemShut {NoStop}%
\bibitem [{\citenamefont {Cardoso}\ and\ \citenamefont
  {Pani}(2019)}]{Cardoso:2019rvt}%
  \BibitemOpen
  \bibfield  {author} {\bibinfo {author} {\bibfnamefont {V.}~\bibnamefont
  {Cardoso}}\ and\ \bibinfo {author} {\bibfnamefont {P.}~\bibnamefont {Pani}},\
  }\href {https://doi.org/10.1007/s41114-019-0020-4} {\bibfield  {journal}
  {\bibinfo  {journal} {Living Rev. Rel.}\ }\textbf {\bibinfo {volume} {22}},\
  \bibinfo {pages} {4} (\bibinfo {year} {2019})},\ \Eprint
  {https://arxiv.org/abs/1904.05363} {arXiv:1904.05363 [gr-qc]} \BibitemShut
  {NoStop}%
\bibitem [{\citenamefont {Claudel}\ \emph {et~al.}(2001)\citenamefont
  {Claudel}, \citenamefont {Virbhadra},\ and\ \citenamefont
  {Ellis}}]{Claudel:2000yi}%
  \BibitemOpen
  \bibfield  {author} {\bibinfo {author} {\bibfnamefont {C.-M.}\ \bibnamefont
  {Claudel}}, \bibinfo {author} {\bibfnamefont {K.~S.}\ \bibnamefont
  {Virbhadra}},\ and\ \bibinfo {author} {\bibfnamefont {G.~F.~R.}\ \bibnamefont
  {Ellis}},\ }\href {https://doi.org/10.1063/1.1308507} {\bibfield  {journal}
  {\bibinfo  {journal} {J. Math. Phys.}\ }\textbf {\bibinfo {volume} {42}},\
  \bibinfo {pages} {818} (\bibinfo {year} {2001})},\ \Eprint
  {https://arxiv.org/abs/gr-qc/0005050} {arXiv:gr-qc/0005050} \BibitemShut
  {NoStop}%
\bibitem [{\citenamefont {Cunha}\ and\ \citenamefont
  {Herdeiro}(2020)}]{Cunha:2020azh}%
  \BibitemOpen
  \bibfield  {author} {\bibinfo {author} {\bibfnamefont {P.~V.~P.}\
  \bibnamefont {Cunha}}\ and\ \bibinfo {author} {\bibfnamefont {C.~A.~R.}\
  \bibnamefont {Herdeiro}},\ }\href
  {https://doi.org/10.1103/PhysRevLett.124.181101} {\bibfield  {journal}
  {\bibinfo  {journal} {Phys. Rev. Lett.}\ }\textbf {\bibinfo {volume} {124}},\
  \bibinfo {pages} {181101} (\bibinfo {year} {2020})},\ \Eprint
  {https://arxiv.org/abs/2003.06445} {arXiv:2003.06445 [gr-qc]} \BibitemShut
  {NoStop}%
\bibitem [{\citenamefont {Guo}\ and\ \citenamefont {Gao}(2021)}]{Guo:2020qwk}%
  \BibitemOpen
  \bibfield  {author} {\bibinfo {author} {\bibfnamefont {M.}~\bibnamefont
  {Guo}}\ and\ \bibinfo {author} {\bibfnamefont {S.}~\bibnamefont {Gao}},\
  }\href {https://doi.org/10.1103/PhysRevD.103.104031} {\bibfield  {journal}
  {\bibinfo  {journal} {Phys. Rev. D}\ }\textbf {\bibinfo {volume} {103}},\
  \bibinfo {pages} {104031} (\bibinfo {year} {2021})},\ \Eprint
  {https://arxiv.org/abs/2011.02211} {arXiv:2011.02211 [gr-qc]} \BibitemShut
  {NoStop}%
\bibitem [{\citenamefont {Ghosh}\ and\ \citenamefont
  {Sarkar}(2021)}]{Ghosh:2021txu}%
  \BibitemOpen
  \bibfield  {author} {\bibinfo {author} {\bibfnamefont {R.}~\bibnamefont
  {Ghosh}}\ and\ \bibinfo {author} {\bibfnamefont {S.}~\bibnamefont {Sarkar}},\
  }\href {https://doi.org/10.1103/PhysRevD.104.044019} {\bibfield  {journal}
  {\bibinfo  {journal} {Phys. Rev. D}\ }\textbf {\bibinfo {volume} {104}},\
  \bibinfo {pages} {044019} (\bibinfo {year} {2021})},\ \Eprint
  {https://arxiv.org/abs/2107.07370} {arXiv:2107.07370 [gr-qc]} \BibitemShut
  {NoStop}%
\bibitem [{\citenamefont {Wei}(2020)}]{Wei:2020rbh}%
  \BibitemOpen
  \bibfield  {author} {\bibinfo {author} {\bibfnamefont {S.-W.}\ \bibnamefont
  {Wei}},\ }\href {https://doi.org/10.1103/PhysRevD.102.064039} {\bibfield
  {journal} {\bibinfo  {journal} {Phys. Rev. D}\ }\textbf {\bibinfo {volume}
  {102}},\ \bibinfo {pages} {064039} (\bibinfo {year} {2020})},\ \Eprint
  {https://arxiv.org/abs/2006.02112} {arXiv:2006.02112 [gr-qc]} \BibitemShut
  {NoStop}%
\bibitem [{\citenamefont {Ye}\ and\ \citenamefont {Wei}(2023)}]{Ye:2023gmk}%
  \BibitemOpen
  \bibfield  {author} {\bibinfo {author} {\bibfnamefont {X.}~\bibnamefont
  {Ye}}\ and\ \bibinfo {author} {\bibfnamefont {S.-W.}\ \bibnamefont {Wei}},\
  }\href@noop {} {\  (\bibinfo {year} {2023})},\ \Eprint
  {https://arxiv.org/abs/2301.04786} {arXiv:2301.04786 [gr-qc]} \BibitemShut
  {NoStop}%
\bibitem [{\citenamefont {Wei}\ and\ \citenamefont {Liu}(2023)}]{Wei:2022mzv}%
  \BibitemOpen
  \bibfield  {author} {\bibinfo {author} {\bibfnamefont {S.-W.}\ \bibnamefont
  {Wei}}\ and\ \bibinfo {author} {\bibfnamefont {Y.-X.}\ \bibnamefont {Liu}},\
  }\href {https://doi.org/10.1103/PhysRevD.107.064006} {\bibfield  {journal}
  {\bibinfo  {journal} {Phys. Rev. D}\ }\textbf {\bibinfo {volume} {107}},\
  \bibinfo {pages} {064006} (\bibinfo {year} {2023})},\ \Eprint
  {https://arxiv.org/abs/2207.08397} {arXiv:2207.08397 [gr-qc]} \BibitemShut
  {NoStop}%
\bibitem [{\citenamefont {Junior}\ \emph {et~al.}(2021)\citenamefont {Junior},
  \citenamefont {Cunha}, \citenamefont {Herdeiro},\ and\ \citenamefont
  {Crispino}}]{Junior:2021dyw}%
  \BibitemOpen
  \bibfield  {author} {\bibinfo {author} {\bibfnamefont {H.~C. D.~L.}\
  \bibnamefont {Junior}}, \bibinfo {author} {\bibfnamefont {P.~V.~P.}\
  \bibnamefont {Cunha}}, \bibinfo {author} {\bibfnamefont {C.~A.~R.}\
  \bibnamefont {Herdeiro}},\ and\ \bibinfo {author} {\bibfnamefont {L.~C.~B.}\
  \bibnamefont {Crispino}},\ }\href
  {https://doi.org/10.1103/PhysRevD.104.044018} {\bibfield  {journal} {\bibinfo
   {journal} {Phys. Rev. D}\ }\textbf {\bibinfo {volume} {104}},\ \bibinfo
  {pages} {044018} (\bibinfo {year} {2021})},\ \Eprint
  {https://arxiv.org/abs/2104.09577} {arXiv:2104.09577 [gr-qc]} \BibitemShut
  {NoStop}%
\bibitem [{\citenamefont {Cunha}\ \emph {et~al.}(2022)\citenamefont {Cunha},
  \citenamefont {Herdeiro},\ and\ \citenamefont {Novo}}]{Cunha:2022nyw}%
  \BibitemOpen
  \bibfield  {author} {\bibinfo {author} {\bibfnamefont {P.~V.~P.}\
  \bibnamefont {Cunha}}, \bibinfo {author} {\bibfnamefont {C.~A.~R.}\
  \bibnamefont {Herdeiro}},\ and\ \bibinfo {author} {\bibfnamefont {J.~a.
  P.~A.}\ \bibnamefont {Novo}},\ }\href
  {https://doi.org/10.1088/1361-6382/ac987e} {\bibfield  {journal} {\bibinfo
  {journal} {Class. Quant. Grav.}\ }\textbf {\bibinfo {volume} {39}},\ \bibinfo
  {pages} {225007} (\bibinfo {year} {2022})},\ \Eprint
  {https://arxiv.org/abs/2207.14506} {arXiv:2207.14506 [gr-qc]} \BibitemShut
  {NoStop}%
\bibitem [{\citenamefont {Bardeen}\ \emph {et~al.}(1972)\citenamefont
  {Bardeen}, \citenamefont {Press},\ and\ \citenamefont
  {Teukolsky}}]{Bardeen:1972fi}%
  \BibitemOpen
  \bibfield  {author} {\bibinfo {author} {\bibfnamefont {J.~M.}\ \bibnamefont
  {Bardeen}}, \bibinfo {author} {\bibfnamefont {W.~H.}\ \bibnamefont {Press}},\
  and\ \bibinfo {author} {\bibfnamefont {S.~A.}\ \bibnamefont {Teukolsky}},\
  }\href {https://doi.org/10.1086/151796} {\bibfield  {journal} {\bibinfo
  {journal} {Astrophys. J.}\ }\textbf {\bibinfo {volume} {178}},\ \bibinfo
  {pages} {347} (\bibinfo {year} {1972})}\BibitemShut {NoStop}%
\bibitem [{\citenamefont {Boonserm}\ \emph {et~al.}(2020)\citenamefont
  {Boonserm}, \citenamefont {Ngampitipan}, \citenamefont {Simpson},\ and\
  \citenamefont {Visser}}]{Boonserm:2019nqq}%
  \BibitemOpen
  \bibfield  {author} {\bibinfo {author} {\bibfnamefont {P.}~\bibnamefont
  {Boonserm}}, \bibinfo {author} {\bibfnamefont {T.}~\bibnamefont
  {Ngampitipan}}, \bibinfo {author} {\bibfnamefont {A.}~\bibnamefont
  {Simpson}},\ and\ \bibinfo {author} {\bibfnamefont {M.}~\bibnamefont
  {Visser}},\ }\href {https://doi.org/10.1103/PhysRevD.101.024050} {\bibfield
  {journal} {\bibinfo  {journal} {Phys. Rev. D}\ }\textbf {\bibinfo {volume}
  {101}},\ \bibinfo {pages} {024050} (\bibinfo {year} {2020})},\ \Eprint
  {https://arxiv.org/abs/1909.06755} {arXiv:1909.06755 [gr-qc]} \BibitemShut
  {NoStop}%
\bibitem [{\citenamefont {Chandrasekhar}(1985)}]{Chandrasekhar:1985kt}%
  \BibitemOpen
  \bibfield  {author} {\bibinfo {author} {\bibfnamefont {S.}~\bibnamefont
  {Chandrasekhar}},\ }\href@noop {} {\emph {\bibinfo {title} {{The mathematical
  theory of black holes}}}}\ (\bibinfo {year} {1985})\BibitemShut {NoStop}%
\bibitem [{\citenamefont {Delgado}\ \emph {et~al.}(2022)\citenamefont
  {Delgado}, \citenamefont {Herdeiro},\ and\ \citenamefont
  {Radu}}]{Delgado:2021jxd}%
  \BibitemOpen
  \bibfield  {author} {\bibinfo {author} {\bibfnamefont {J.~F.~M.}\
  \bibnamefont {Delgado}}, \bibinfo {author} {\bibfnamefont {C.~A.~R.}\
  \bibnamefont {Herdeiro}},\ and\ \bibinfo {author} {\bibfnamefont
  {E.}~\bibnamefont {Radu}},\ }\href
  {https://doi.org/10.1103/PhysRevD.105.064026} {\bibfield  {journal} {\bibinfo
   {journal} {Phys. Rev. D}\ }\textbf {\bibinfo {volume} {105}},\ \bibinfo
  {pages} {064026} (\bibinfo {year} {2022})},\ \Eprint
  {https://arxiv.org/abs/2107.03404} {arXiv:2107.03404 [gr-qc]} \BibitemShut
  {NoStop}%
\bibitem [{\citenamefont {Collodel}\ \emph {et~al.}(2018)\citenamefont
  {Collodel}, \citenamefont {Kleihaus},\ and\ \citenamefont
  {Kunz}}]{Collodel:2017end}%
  \BibitemOpen
  \bibfield  {author} {\bibinfo {author} {\bibfnamefont {L.~G.}\ \bibnamefont
  {Collodel}}, \bibinfo {author} {\bibfnamefont {B.}~\bibnamefont {Kleihaus}},\
  and\ \bibinfo {author} {\bibfnamefont {J.}~\bibnamefont {Kunz}},\ }\href
  {https://doi.org/10.1103/PhysRevLett.120.201103} {\bibfield  {journal}
  {\bibinfo  {journal} {Phys. Rev. Lett.}\ }\textbf {\bibinfo {volume} {120}},\
  \bibinfo {pages} {201103} (\bibinfo {year} {2018})},\ \Eprint
  {https://arxiv.org/abs/1711.05191} {arXiv:1711.05191 [gr-qc]} \BibitemShut
  {NoStop}%
\end{thebibliography}
\end{document}